\documentclass[aps,prd,twocolumn,groupedaddress,showpacs]{revtex4-1}

\usepackage{graphicx}% Include figure files
\usepackage{dcolumn}% Align table columns on decimal point
\usepackage{bm}% bold math

\begin{document}

\title{Probing the photon flux in the diffractive quarkonium photoproduction at LHC}

\author{V. P. Gon\c calves}

\email[]{barros@ufpel.edu.br}

\affiliation{Instituto de F\'{\i}sica e Matem\'atica, Universidade Federal de
Pelotas\\
Caixa Postal 354, CEP 96010-090, Pelotas, RS, Brazil}

\author{G. Gil da Silveira}

\email[]{gustavo.silveira@cern.ch}

\affiliation{Instituto de F\'{\i}sica e Matem\'atica, Universidade Federal de
Pelotas\\
Caixa Postal 354, CEP 96010-090, Pelotas, RS, Brazil}

\begin{abstract}
In this paper we propose the study of the diffractive quarkonium photoproduction in $pp$ collisions at LHC energies  to probe the photon flux associated to a ultrarelativistic proton. The total photon distribution is expected to be given in terms of the elastic and inelastic components. Distinctly from the elastic photon component which can be determined from the elastic form factors, the magnitude of the inelastic component still is an open question. We consider the current parametrizations for the photon distribution of a proton and estimate the 
 rapidity distributions and total cross sections for the 
production of $J/\Psi$, $\Psi^{\prime}$ and $\Upsilon$ at LHC energies. We demonstrate that the predictions associated to the  inelastic contribution are of the same order than those associated to the elastic one. Our results implies that   a dedicated experimental analysis of diffractive quarkonium photoproduction with the tagging of the two protons in the final state  can be useful to constrain the magnitude of the inelastic component of the photon distribution.
\end{abstract}


\maketitle

\section{Introduction}

\begin{figure}
\begin{tabular}{cc}
\includegraphics[scale=0.25]{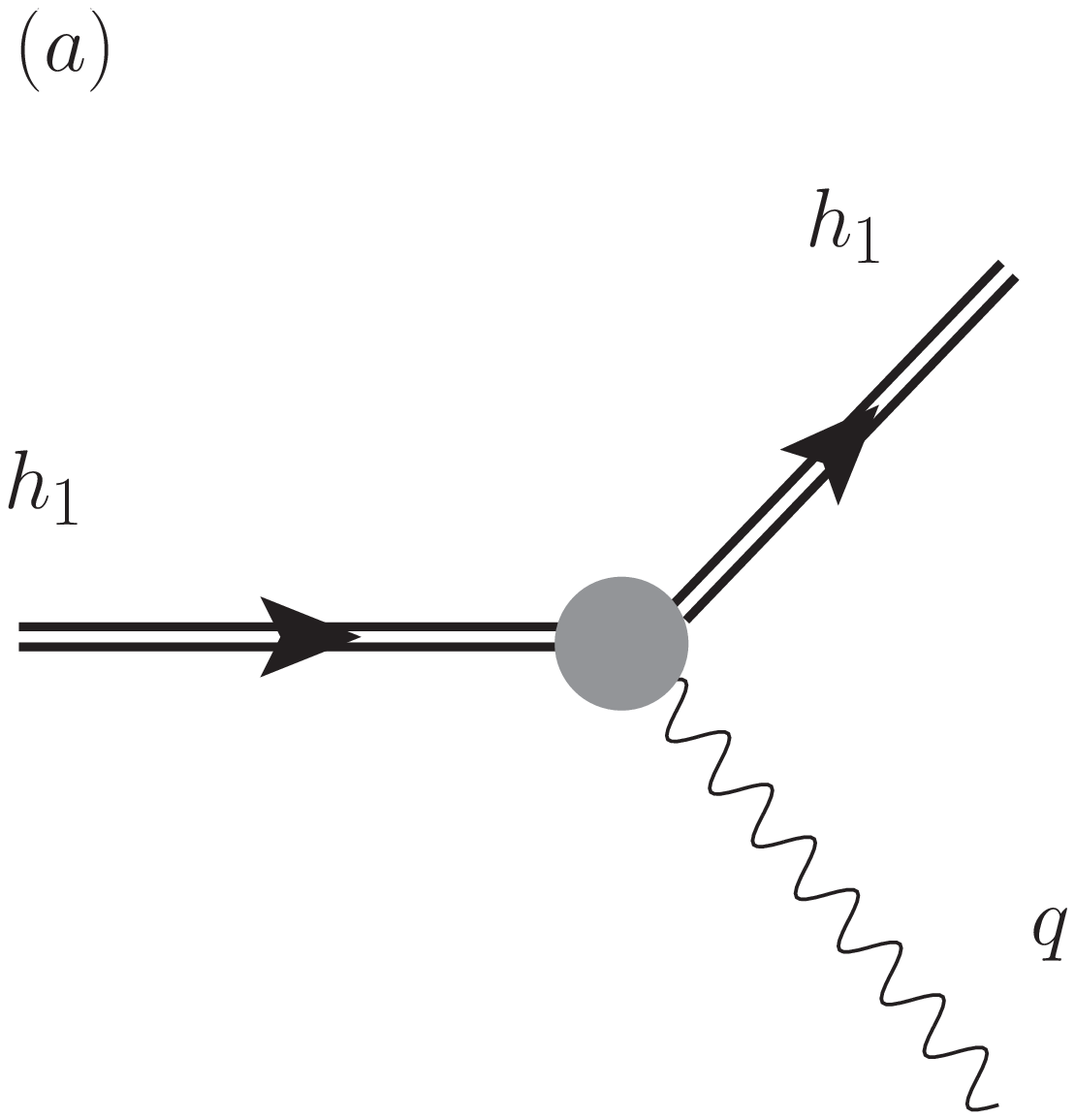} &
\includegraphics[scale=0.25]{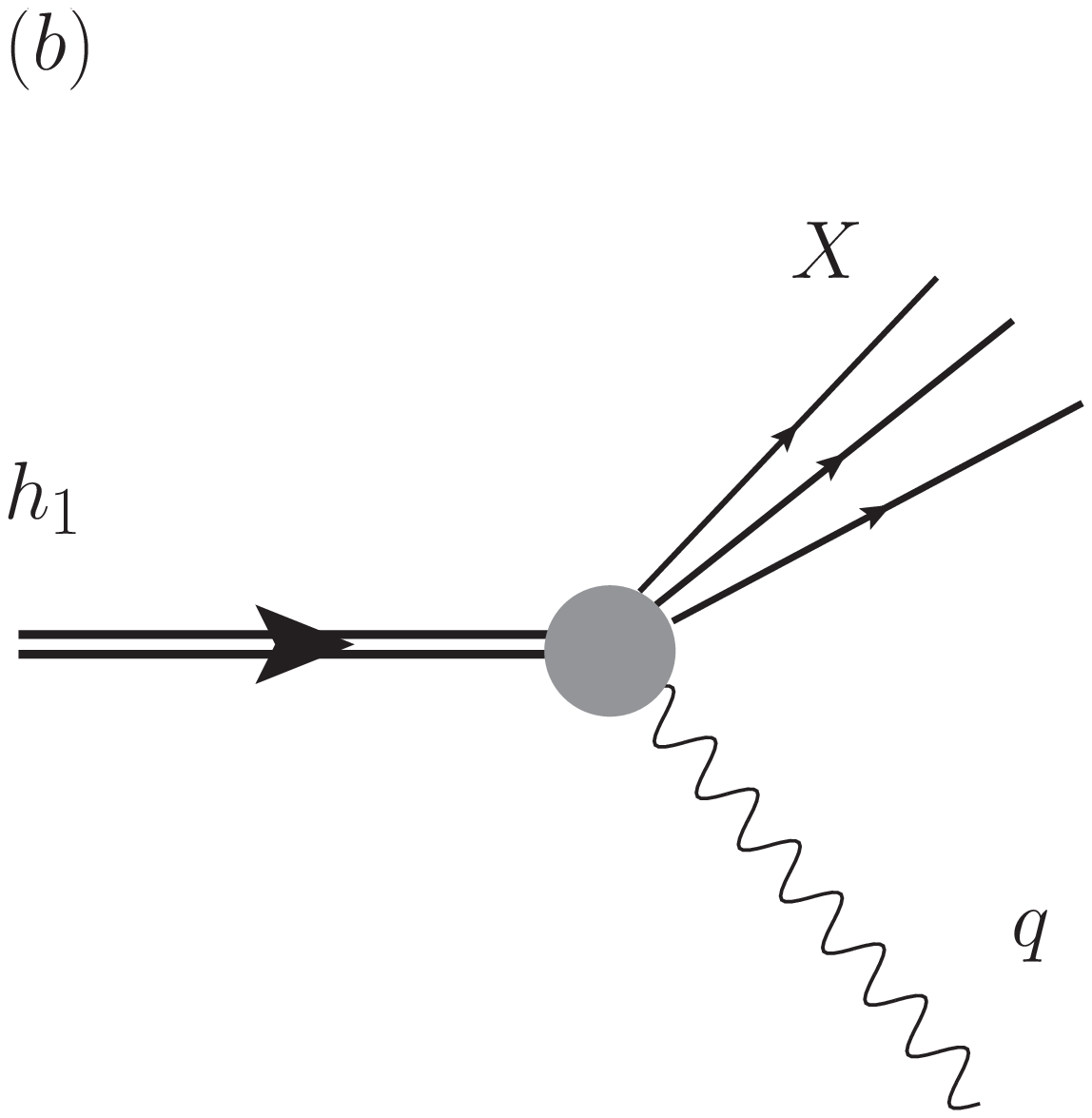}
\end{tabular}
\caption{ (a) Elastic and (b) Inelastic components of the equivalent photon distributions of a hadron.}
\label{diagrama_flux}
\end{figure}

 In recent  years a series of experimental results at RHIC \cite{star,phenix}, Tevatron \cite{cdf} and LHC \cite{alice, alice2,lhcb,lhcb2}  demonstrated that the study of photon - induced interactions in hadronic colliders is feasible and can be used to probe e.g. the nuclear gluon  distribution \cite{gluon,gluon2,Guzey}, the dynamics of the strong interactions \cite{vicmag_mesons1,outros_vicmag_mesons,vicmag_update,motyka_watt,Lappi,griep} and the mechanism  of quarkonium production \cite{Schafer,mairon1,mairon2,cisek,bruno1,bruno2}. It has stimulated the improvement of the theoretical description of these processes as well as the proposal of new forward detectors to be installed in the LHC.
The basic idea in the photon-induced processes is that
 a ultra relativistic charged hadron (proton or nuclei)
 give rise to strong electromagnetic fields, such that the photon stemming from the electromagnetic field
of one of the two colliding hadrons can interact with one photon of
the other hadron (photon - photon process) or can interact directly with the other hadron (photon - hadron
process) \cite{upc}. In these processes the total cross section  can be factorized in
terms of the equivalent flux of photons into the hadron projectile and the photon-photon or photon-target production cross section.  Consequently, a basic ingredient in the analysis of the photon - induced processes is the description of the  equivalent photon distribution of the hadron. The equivalent photon approximation of a charged  pointlike fermion was formulated  many years ago by Fermi \cite{Fermi} and developed by Williams \cite{Williams} and Weizsacker \cite{Weizsacker}. In contrast, the calculation of the photon distribution of the hadrons still is a subject of debate, due to the fact that they are not point like particles. In this case it is necessary to distinguish between the  elastic and inelastic components (See Fig. \ref{diagrama_flux}).  The elastic component [Fig. \ref{diagrama_flux} (a)] can be estimated analysing the transition $h \rightarrow \gamma h$ taking into account the effects of the hadronic form factors, with the hadron remaining intact in the final state \cite{kniehl}. In contrast, the inelastic contribution [Fig. \ref{diagrama_flux} (b)] is associated to the transition $h \rightarrow \gamma X$, with $X \neq h$, and  can be estimated taking into account the partonic structure of the hadrons, which can be a source of photons (See, e.g. Refs. \cite{epa,rujula,drees_godbole,pisano,mrstqed,nnpdf,cteqqed,martin_ryskin}). Therefore the total photon distribution of a hadron is given by
\begin{eqnarray}
\gamma(x,\mu^2) = \gamma_{el} (x) + \gamma_{inel} (x,\mu^2) 
\label{total}
\end{eqnarray}
where $x$ is the fraction of the hadron energy carried by the photon and $\mu$ has to be identified with a momentum scale of the photon - induced process. It is important to emphasize that while $\gamma_{el}$ is proportional to squared charge of the hadron ($Z^2$),  due to the coherent action of all protons in the nucleus, $\gamma_{inel}$ is proportional to the mass number $A$. Consequently, for a heavy nuclei, the total photon distribution is determined by its elastic component. In contrast, for a proton, both components contribute equally and should be taking into account in the study of photon - induced processes. Currently, the description of the inelastic component still is an open question, with the predictions for its $x$ dependence being largely distinct, as we will demonstrate in the next section. Our goal in this paper is twofold: (a) analyse the impact of the inelastic component in the diffractive photoproduction of vector mesons ($V = J/\Psi, \Psi^{\prime}$ and $\Upsilon$) in $pp$ collisions at LHC, and (b) verify if a more detailed  analysis  of this process can be used to constrain the inelastic component of the photon distribution.

This paper is organized as follows. In the next section we briefly review the description of the elastic and inelastic components of the photon distribution of a nucleon. In particular, we present a comparison between the current parametrizations for the photon distribution. In Section \ref{dif} we discuss the diffractive photoproduction in $pp$ collisions and the main input used in our calculations. In Section \ref{res} we present our results for the $J/\Psi, \Psi^{\prime}$ and $\Upsilon$ production considering the different models for the inelastic component of the photon distribution and compare with those obtained considering the elastic component. Finally, in Section \ref{sum}, we summarize our main conclusions.

\section{The equivalent photon distribution}
The concept of the photon content of a charged fermion is based on the equivalent photon approximation (EPA) \cite{epa,upc}, which implies that the photon distribution of a nucleon consist of two parts: the elastic and inelastic components. A detailed derivation of the  elastic photon distribution of a nucleon was presented in Ref. \cite{kniehl} which can be written as
\begin{widetext}
\begin{eqnarray}
\gamma_{el} (x) = - \frac{\alpha}{2\pi} \int_{-\infty}^{-\frac{m^2x^2}{1-x}} \frac{dt}{t}\left\{\left[2\left(\frac{1}{x}-1\right) + \frac{2m^2x}{t}\right]H_1(t) + xG_M^2(t)\right\}\,\,,
\label{elastic}
\end{eqnarray}
\end{widetext}
 where $t = q^2$ is the momentum transfer squared of the photon, 
 \begin{eqnarray}
 H_1(t) \equiv \frac{G_E^2(t) + \tau G_M^2(t) }{1 + \tau}
 \end{eqnarray}
 with $\tau \equiv -t/m^2$, $m$ being the nucleon mass, and where $G_E$ and $G_M$ are the Sachs elastic form factors.
 Although an analytical expression for the elastic component is presented in Ref. \cite{kniehl}, it is common to found in the literature the study of photon - induced processes considering an approximated expression proposed in Ref. \cite{dz}, which can be obtained from Eq. (\ref{elastic}) by disregarding the contribution of the magnetic dipole moment and the corresponding magnetic form factor. As demonstrated in Refs. \cite{vic_wer_daniel} the difference between the full and the approximated expression is smaller than 5\% at low-$x$. Consequently, in what follows we will use the expression proposed in Ref. \cite{dz}, where the elastic photon distribution is given by
 \begin{eqnarray}
\gamma_{el}(x)&=&\frac{\alpha}{\pi}\left(\frac{1-x+0.5x^{2}}{x}\right) \times \nonumber \\
&\cdot& \left[\ln(\Omega)
-\frac{11}{6}+\frac{3}{\Omega}-\frac{3}{2\Omega^{2}}+\frac{1}{3\Omega^{3}}\right]\,\,,
\label{dz}
\end{eqnarray}
where $\Omega=1+(0.71 \mathrm{GeV}^{2})/Q_{min}^{2}$ and 
 $Q^2_{min} \approx (x m)^2/(1-x)$.

\begin{figure}[t]
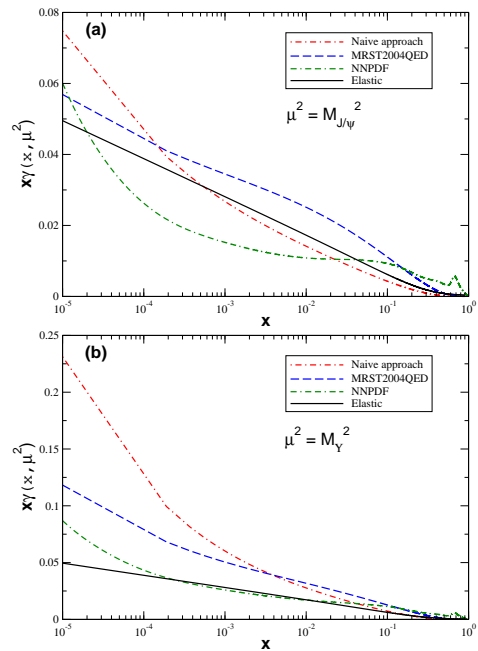

%\begin{tabular}{cc}
\includegraphics[scale=0.25]{fotons_q2_jpsi.eps} \\
\includegraphics[scale=0.25]{fotons_q2_ups.eps}
%\end{tabular}
\caption{(Color online) Comparison between the different models for the inelastic component of the photon distribution for two different values of the hard scale $\mu^2$: (a) $\mu^2 = M_{J/\Psi}^2$ and (b) $\mu^2 = M_{\Upsilon}^2$. The elastic component is presented for comparison. }
\label{ineflux}
\end{figure}

 On the other hand, there are different models for the  contribution of the inelastic component of the photon distribution of a nucleon. In Ref. \cite{drees_godbole}, a naive approach to the photon flux was proposed, with the photon distribution in the proton being given by a convolution of the distribution of quarks in the proton and the distribution of photons in the quarks as follows
 \begin{eqnarray}
 \gamma_{inel}(x,\mu^ 2) = \sum_q \int_x^1 \frac{dx_q}{x_q} f_q(x_q,\mu^2) e_q^2 f_{\gamma/q}(\frac{x}{x_q},Q_1^2,Q_2^2) \,\,, \label{naive}
\end{eqnarray}   
 where the sum runs over all quark and antiquark flavours and the flux of photons in a quark $f_{\gamma/q}$ is given by
 \begin{eqnarray}
 f_{\gamma/q} (z) = \frac{\alpha}{2\pi} \frac{1+(1-z)^2}{z} \log \frac{Q_1^2}{Q_2^2} \,\,,
\end{eqnarray}  
with $Q_1^2$ being assumed to be the maximum value of the momentum transfer in the process and $Q_2^2$ is assumed to be equal to 1 GeV$^2$ in order to the parton model to be applicable.  Recently, different groups have studied the modification of  the Dokshitzer - Gribov - Lipatov - Altarelli - Parisi equations for the quark and gluon distributions by the inclusion of QED contributions and have performed global parton analysis of deep inelastic and related hard-scattering data \cite{mrstqed,nnpdf,cteqqed,martin_ryskin}. 
Basically, the DGLAP equations and the momentum sum  rule are modified considering the presence of the  photon as an additional point-like parton in the nucleon. The parametrizations for the photon distribution currently available in the literature \cite{mrstqed,nnpdf} differ in the approach for the initial condition for the photon distribution. While the MRST group assume that  $\gamma_{inel} (x,Q_0^2)$ is given by a expression similar to Eq. (\ref{naive}), the NNPDF group parametrize the  input photon PDF and attempt to determine the parameters from the global data.  The preliminary CTEQ analysis presented in Ref. \cite{cteqqed} assume a similar theoretical form for  $\gamma_{inel} (x,Q_0^2)$ to that proposed by the MRST group, but with an arbitrary normalization parameter, which is expressed as the momentum fraction carried by the photon. More recently, a distinct approach for the initial condition for the evolution of the photon distribution was proposed in Ref. \cite{martin_ryskin}, where the authors have proposed  that the starting distribution for the photon PDF
should be the total photon distribution, i.e. by the sum of the elastic and inelastic components as given in Eq. (\ref{total}). The main motivation of this approach is the reduced uncertainty in the input photon PDF, since the major part of the distribution is given by the elastic component, which is well  known. As a consequence of this assumption, the elastic component is dominant also at large values of the hard scale $\mu^2$ (See Fig. 5 in \cite{martin_ryskin}).
Unfortunately, the current data is not sufficient accurate to precisely determine the initial condition. Thus the current predictions for the inelastic photon component strongly differ in its $x$ dependence. In what follows we  will consider  the MRST2004QED and NNPDF parametrizations, since only these two are currently available to  public use. 
In Fig. \ref{ineflux} we present the predictions of the MRST2004QED and NNPDF parametrizations for the inelastic photon distribution considering two different values for the hard scale $\mu^2$. For comparison the predictions of the Naive approach [Eq. (\ref{naive})] and the elastic component [Eq. (\ref{dz})] are also presented. While the elastic component is independent of the hard scale $\mu^2$, the inelastic component is strongly dependent, increasing at larger values of $\mu^2$. Moreover, all models predict that the inelastic contribution is dominant at very small values of $x$. However, as demonstrated in the figure, the $x$-dependence of the inelastic parametrizations is very distinct. This result motivates the analysis of observables which are strongly dependent on the photon flux.

\section{The diffractive quarkonium photoproduction in $pp$ collisions}
\label{dif}
The basic idea for the description of the diffractive quarkonium photoproduction in $pp$ collisions is that the  total
cross section  can be factorized in terms of the equivalent flux of photons of the hadron projectile and  the  photon-target production cross section  as follows \cite{upc}
\begin{eqnarray}
\sigma (h_1 h_2 \rightarrow  p \otimes V \otimes h_3 ) =  \sum_{i=1,2} \int dY \frac{d\sigma_i}{dY}\,,
\label{sighh}
\end{eqnarray}
where  $h_1 = h_2 = p$, $\otimes$ represents a rapidity gap in the final state and $h_3 = p$ or $X$ depending if the 
incident proton which emits the photon remains intact or dissociate.  Moreover,  ${d\sigma_i}/{dY}$ is the rapidity distribution for the photon-target interaction induced by the hadron $h_i$ ($i =1,2$), which can be expressed as 
\begin{eqnarray}
\frac{d\sigma_i}{dY} = x\gamma_i (x,\mu^2)\,\sigma_{\gamma h_j \rightarrow V h_j} (W_{\gamma h_j}^2) \,\,\,\,\,\,(i\neq j)\,,
\label{rapdis}
\end{eqnarray}
where $\gamma_i$ is the equivalent photon flux associated to the hadron $i$,     $W_{\gamma h}^2=2\,\omega \sqrt{s_{\mathrm{NN}}}$ is the c.m.s energy squared of the
photon - hadron system, $\omega$ is the photon energy  and ${s_{\mathrm{NN}}}$ is  the  c.m.s energy squared of the
hadron-hadron system.

\begin{figure}[t]
%\begin{tabular}{cc}
\includegraphics[scale=0.25]{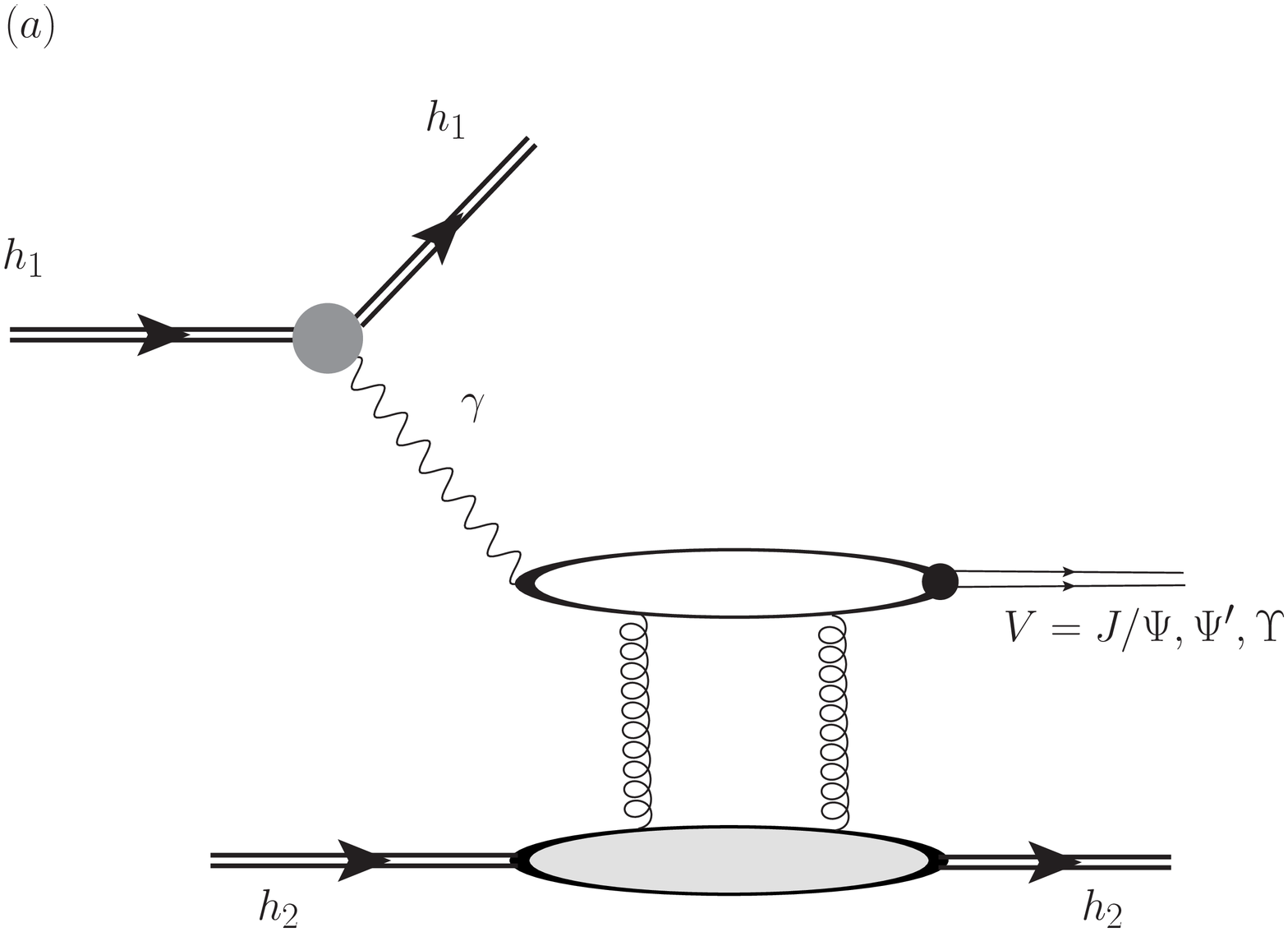} \\
\includegraphics[scale=0.25]{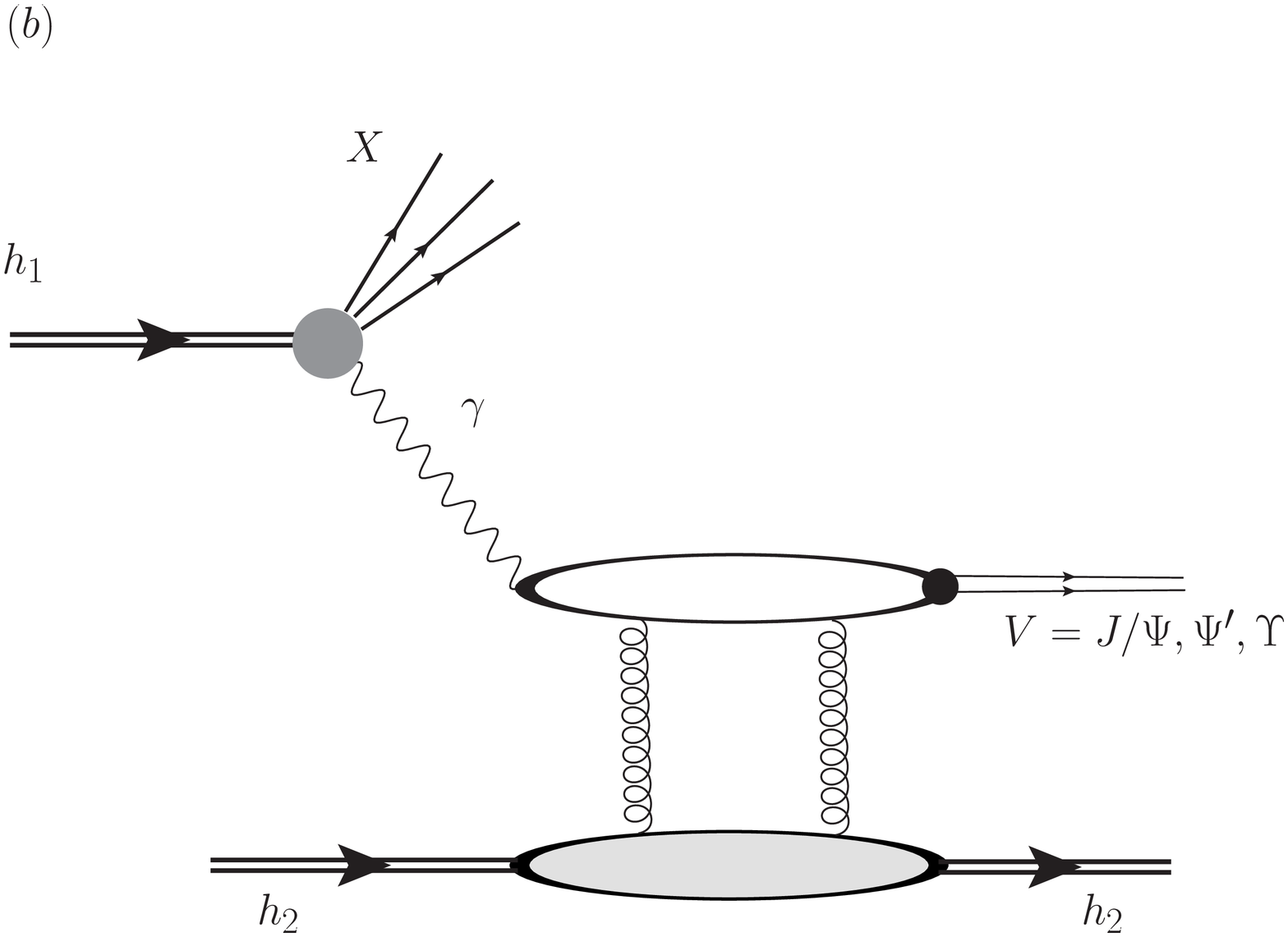}
%\end{tabular}
\caption{ Diffractive quarkonium photoproduction in  hadronic collisions associated to the (a) elastic and (b) inelastic components of the photon distribution of the hadron. }
\label{quarkonium}
\end{figure}

Since the pioneering studies \cite{klein_prc,gluon,strikman} on diffractive vector meson production in ultra peripheral heavy ion collisions (UPHIC) about fourteen years ago, a large number of papers on the subject has been published considering  several improvements in the theoretical description \cite{outros_klein,vicmag_mesons1,outros_vicmag_mesons,outros_frankfurt,Schafer,vicmag_update,gluon2,motyka_watt,Lappi,griep,Guzey,Martin,cisek,bruno1, bruno2}  and experimental analysis \cite{cdf,star,phenix,alice,alice2,lhcb}. However, such studies have only considered the process in which the hadron which emits the photon remain intact represented in Fig. \ref{quarkonium} (a). In other words, these studies have assumed that the total photon distribution is dominated by the elastic component and that the hadron which emits the photon remains intact. As discussed in the Introduction, such  approximation is reasonable for a nuclear projectile. On the other hand, for a proton projectile, the magnitude of the contribution associated to the inelastic component of the photon distribution, where the proton which emits the photon dissociates, represented in Fig. \ref{quarkonium} (b), remains an open question. This is the main goal of the next section. Before it, we need to specify the  cross section for the diffractive photoproduction of a vector meson  ($\sigma_{\gamma p \rightarrow V p}$).
In recent years such process was studied by several theoretical groups considering different formalisms and underlying assumptions, with its predictions in reasonable agreement with the current experimental data. In order to obtain predictions of the inelastic contribution which are not dependent on the choice of the model used to estimate $\sigma_{\gamma p \rightarrow V p}$,  we will assume in what follows the following form obtained in the H1 analyses \cite{h1}
\begin{eqnarray}
\sigma_{\gamma p \rightarrow J/\Psi p} (W_{\gamma p}) = N \left(\frac{W_{\gamma p}}{90 \mbox{GeV}}\right)^{\lambda} \,\,,\nonumber \\
\sigma_{\gamma p \rightarrow \Psi^{\prime} p} (W_{\gamma p}) = 0.166 N \left(\frac{W_{\gamma p}}{90 \mbox{GeV}}\right)^{\lambda} \,\,,
\label{h1fit}
\end{eqnarray}
where $N = 81 \pm 3$ nb and $\lambda  = 0.67 \pm 0.03$. Moreover, for the $\Upsilon$ production, we will assume that 
 $\sigma_{\gamma p \rightarrow \Upsilon p} = (0.12 \mbox{pb})(W_{\gamma p}/W_0)^{1.6}$ with $W_0 = 1$ GeV, as given in Ref. \cite{motyka_watt}.

\section{Results}
\label{res}

Lets initially calculate the rapidity distribution and total cross section for the diffractive quarkonium photoproduction in $pp$ collisions at LHC energies. 
The distribution on rapidity $Y$ of a vector meson $V$ ($= J/\Psi, \Psi^{\prime}, \Upsilon$)  of mass $M_V$ in the final state can be directly computed from Eq. (\ref{sighh}), by using its  relation with the photon energy $\omega$, i.e. $Y\propto \ln \, ( \omega/M_V)$.  Explicitly, the rapidity distribution is written down as, 
\begin{widetext}
\begin{eqnarray}
\frac{d\sigma \,\left[h_1 h_2 \rightarrow   p \otimes V \otimes h_3 \right]}{dY} = \left[x\gamma_{h_1} (x,\mu^2)\,\sigma_{\gamma h_2 \rightarrow  V \otimes h_2}\left(W_{\gamma h_2}^2 \right)\right]_{\omega_L} + \left[x\gamma_{h_2} (x,\mu^2)\,\sigma_{\gamma h_1 \rightarrow V  \otimes h_1}\left(W_{\gamma h_1}^2 \right)\right]_{\omega_R}\,
\label{dsigdy2}
\end{eqnarray}
\end{widetext}
where  $\omega_L \, (\propto e^{-Y})$ and $\omega_R \, (\propto e^{Y})$ denote photons from the $h_1$ and $h_2$ hadrons, respectively.  Moreover,  $h_3 = p$ or $X$ depending whether the 
incident proton which emits the photon remains intact or dissociate, respectively. As the photon fluxes have support at small values of $\omega$ (low-$x$), decreasing  at large $\omega$ (high-$x$), the first term on the right-hand side of the Eq. (\ref{dsigdy2}) peaks at positive rapidities while the second term peaks at negative rapidities. Consequently,  the study of the rapidity distribution can be used to constrain  the equivalent photon distribution. Moreover, the total rapidity distributions for $pp$  collisions will be symmetric about midrapidity ($Y=0$).

In order to estimate the diffractive quarkonium photoproduction associated to the inelastic component of the photon distribution we should to specify the hard scale $\mu^2$. As can be verified in the literature, the choice of this scale is a bit ambiguous \cite{drees_godbole,zerwas,peterson,mrstqed,nnpdf,antoni}. In general it is assumed that it is related to the center-of-mass energy of the photon -  induced subprocess or to a hard scale in the final state. Following previous analysis that demonstrate that the mass of the vector meson can be considered a hard scale which justifies a perturbative calculation of its photoproduction (See, e.g. Ref. \cite{rmp}), in what follows we will assume that $\mu^2 = M_V^2$. It is important to emphasize that larger values of the hard scale increase our predictions, since the magnitude of the inelastic photon distribution is amplified by the DGLAP evolution. Moreover, in order to estimate the inelastic component using the Naive approach given by Eq. (\ref{naive}) we will assume that $\mu^2 = Q_1^2 = M_V^2$ and that the parton distributions are given by the MRST 2001 leading order parametrization \cite{mrst}.

\begin{figure}[t]
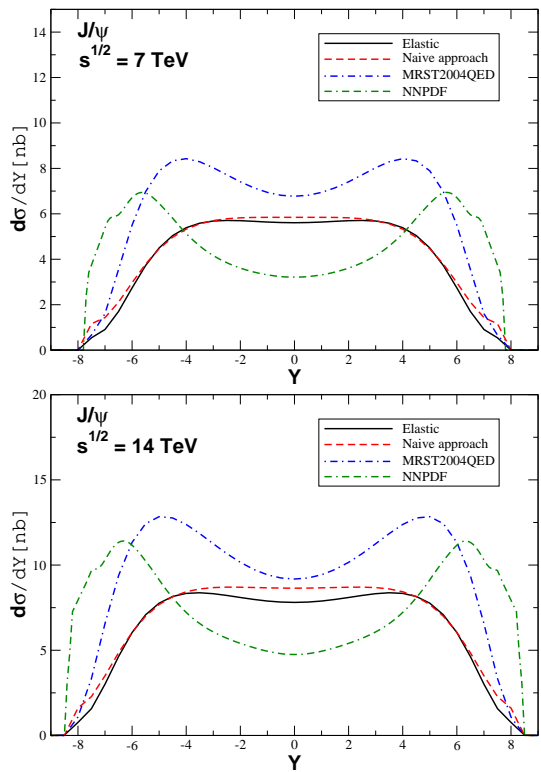

\includegraphics[scale=0.3]{jpsi_7000.eps} \\
\includegraphics[scale=0.3]{jpsi_14000.eps}
\caption{(Color online) Rapidity distribution for the diffractive $J/\Psi$ photoproduction  in $pp$ collisions at  LHC ($\sqrt{s} = 7$ and 14 TeV) considering different models for the inelastic component of the photon distribution.  The predictions associated to the elastic component are also presented for comparison.}
\label{jpsi}
\end{figure}

In Fig. \ref{jpsi} we present our predictions for the rapidity distribution of the diffractive $J/\Psi$ photoproduction in $pp$ collisions at $\sqrt{s} = 7$ and 14 TeV. We consider three different models for the inelastic component of the photon distribution and also present the predictions associated to the elastic contribution  for comparison. We obtain that at midrapidities the NNPDF predictions are a factor $\approx 1.5$ smaller than the elastic one. On the other hand, the MRST2004QED predictions are larger than elastic one, with the Naive predictions being of the same order of the elastic one. Our results indicate that inelastic predictions obtained using the MRST2004QED parametrization are larger than the elastic one in the full rapidity range. In contrast, the NNPDF parametrization implies that the inelastic contribution is  larger than the elastic one at large rapidities. These behaviours are directly related to the $x$-dependence of the inelastic  component of the photon distribution for $\mu^2 = M_{J/\Psi}^2$ presented in Fig. \ref{ineflux} (a), since at large rapidities we are probing larger values of $x$.

\begin{figure}[t]
\includegraphics[scale=0.3]{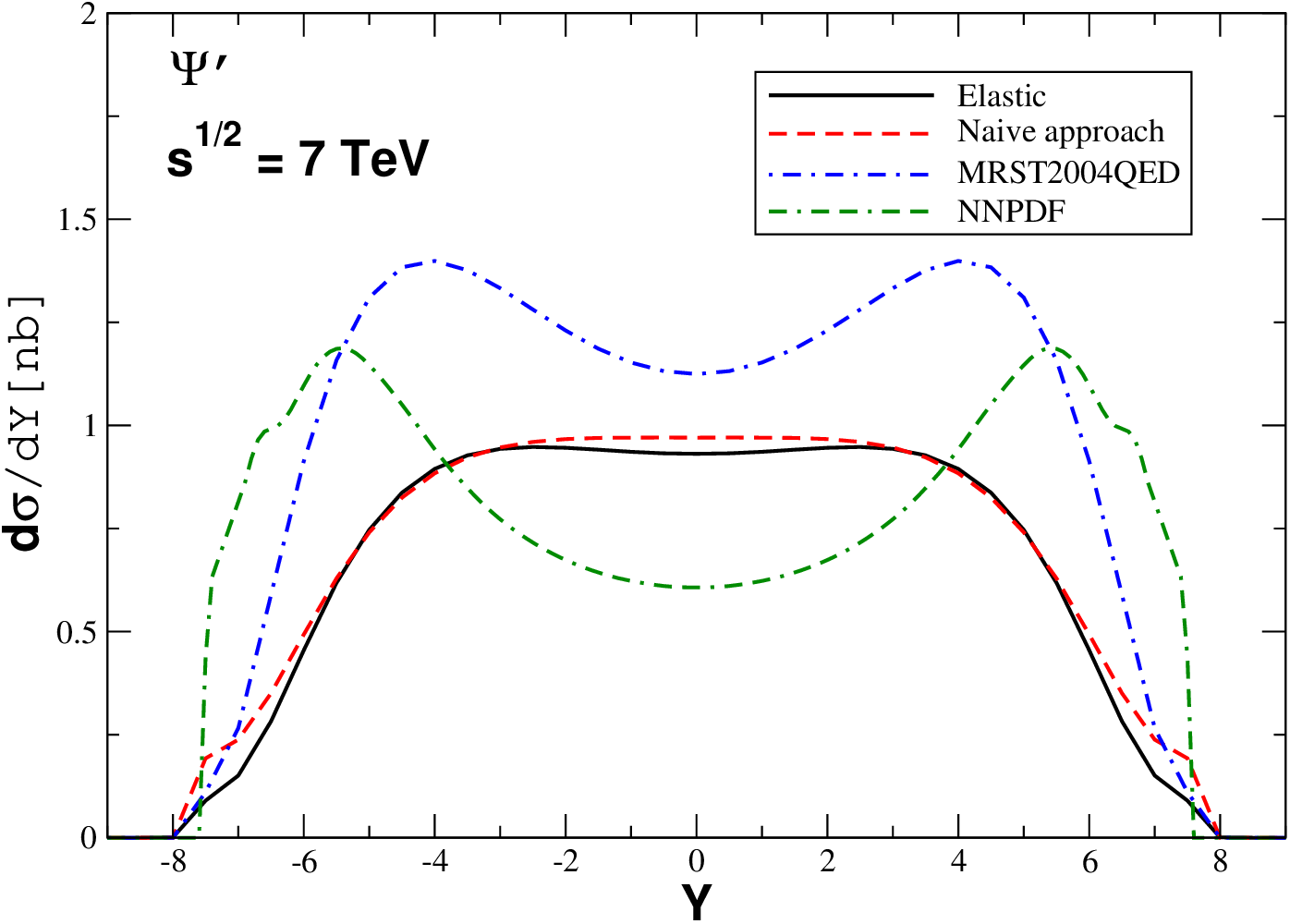} \\
\includegraphics[scale=0.3]{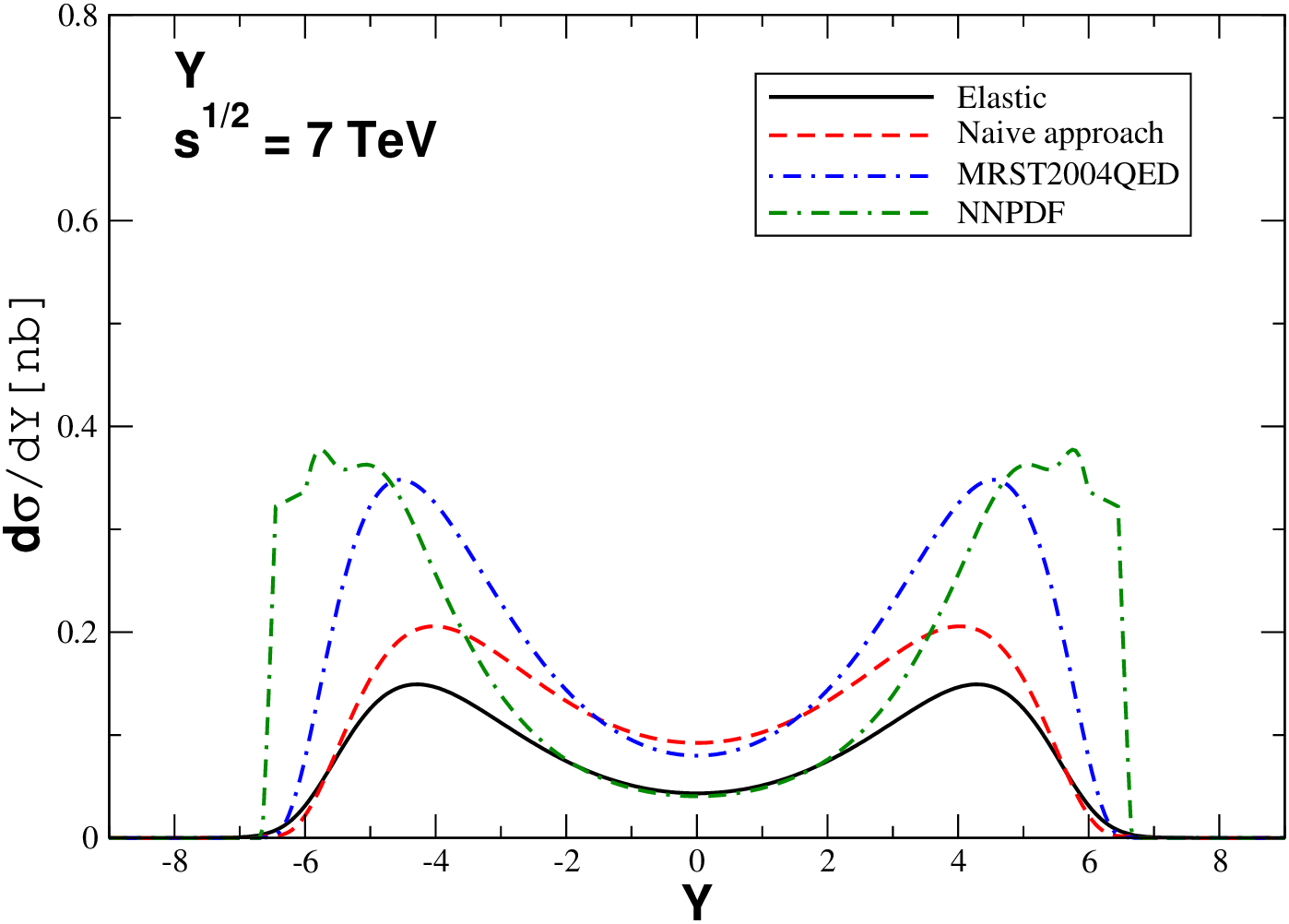}
\caption{(Color online) Rapidity distribution for the diffractive photoproduction of $\Psi^{\prime}$ and $\Upsilon$  in $pp$ collisions at  LHC ($\sqrt{s} = 7$ TeV) considering different models for the inelastic component of the photon distribution.  The predictions associated to the elastic component are also presented for comparison.}
\label{outros_mesons}
\end{figure}

In Fig. \ref{outros_mesons} we present our predictions for the $\Psi^{\prime}$ and $\Upsilon$ photoproduction in $pp$ collisions at $\sqrt{s} = 7$ TeV. For the $\Psi^{\prime}$ production we obtain that the behaviour of the rapidity distributions are very similar to those obtained for the $J/\Psi$ case, which is expected since the energy dependence of the photon - proton cross sections predicted by the H1 parametrizations are the same [See Eq. (\ref{h1fit})], differing only in the normalization. In contrast, for the $\Upsilon$ production, we now obtain that at midrapidities the inelastic NNPDF prediction is of the same order than the elastic one. Moreover, we obtain that the inelastic predictions dominate at large rapidities. As in the $J/\Psi$ case, these behaviours are directly related to the $x$-dependence of the inelastic  component of the photon distribution $\mu^2 = M_{\Upsilon}^2$ presented in Fig. \ref{ineflux} (b).

In Table \ref{tab1} we present our predictions for the total cross sections for the different final states discussed above. In particular, we present  the  predictions associated to the three different models for the inelastic component of the photon distribution. The predictions associated to the elastic component are also presented for comparison.
As expected from our predictions for the rapidity distributions, our results indicate that for the three models of the  inelastic component considered the associated cross sections are larger than the elastic one. These predictions are very distinct and allow, in principle, to discriminate between the different models for the inelastic component of the photon distribution. Moreover, if future experimental results indicate a very small fraction of inelastic processes, it can be considered an evidence that the more adequate approach for the treatment of the photon distribution of the proton is that proposed in Ref. \cite{martin_ryskin}. In order to perfom such study a dedicated experimental analysis is necessary to separate the inelastic and elastic processes. Basically, it is fundamental to tag the two protons into the final state to separate the elastic contribution. In principle, the products of the proton dissociation in the inelastic processes will travel essentially along the beam pipe. Consequently, both processes will be characterized by two rapidity gaps. Therefore, the presence of forward detectors will be essential to characterize the events.

\begin{table}
\begin{center}
\begin{tabular} {|c|c|c|c|c|}
\hline
\hline
Vector Meson &  Naive & MRST2004 & NNPDF & Elastic \\
\hline
\hline
$J/\Psi$ & 69.50 & 98.70 & 73.70 & 66.90  \\
\hline
$\Psi^{\prime}$ & 11.50  & 16.40  & 12.80 & 11.10  \\
\hline
$\Upsilon$  & 1.70  & 2.40 & 2.50 & 1.10 \\ 
\hline
\hline
\end{tabular}
\end{center}
\caption{Total cross sections for the photoproduction of different final states in $pp$ collisions at  LHC ($\sqrt{s} = 7$ TeV)  considering different models for the inelastic component of the photon distribution. For comparison the predictions considering the elastic component are also presented. Values in nb.}
\label{tab1}
\end{table}

\section{Summary} 
\label{sum}

During the last two decades a rich phenomenology of photon - induced processes in hadronic colliders  has emerged. However, several questions still remain to be answered. In particular, the treatment of the photon flux associated to a ultrarelativistic proton still is an open question. In this paper we have proposed, for the first time, the study of the diffractive quarkonium photoproduction  in $pp$ collisions at  LHC energies as a probe of the photon distribution of a proton. This distribution is expected to be characterized by  elastic and inelastic components, which are associated  to the coherent or incoherent emission of the photon from the proton, with the proton remaining intact or dissociating, respectively.
Currently, several groups have proposed distinct approaches for the treatment of the inelastic contribution and its evolution. In this paper we have considered three different models and estimated the diffractive photoproduction of  $J/\Psi$, $\Psi^{\prime}$ and $\Upsilon$ in $pp$ collisions. Our results indicated that, for the models considered, the contribution of the inelastic processes is of the same order or larger than the elastic one, with the predictions for the rapidity distributions being largely different, which makes the experimental discrimination feasible, with the detection of the two protons into the final state being indispensable to separate the inelastic and elastic events. Finally, if the contribution of the inelastic events is probed to be very small, it can be interpreted as a signature of a different approach for the treatment of the  photon distribution \cite{martin_ryskin}.

%\begin{widetext}

%\end{widetext}

\section*{Acknowledgements}
This work has been supported by CNPq, CAPES and FAPERGS, Brazil.

\end{document}